\documentclass[doublecolumns,10pt]{IEEEtran}
\usepackage{amsfonts,epsfig,cite,array,multirow,graphicx,amsmath,ltablex,tabularx,setspace,arydshln,amssymb,multirow}
\usepackage[numbers,sort&compress]{natbib}
\usepackage{verbatim}
\usepackage{graphicx}
\usepackage{epstopdf}
\usepackage{graphicx}
\usepackage{mathtools}

\usepackage{algorithm}
\usepackage{algpseudocode}

\begin{document}

\title{
{Hybrid NOMA for Future Radio Access: Design, Potentials  and Limitations}\\
}
\author{
	\IEEEauthorblockN{Kuntal Deka$^{1}$  and  Sanjeev Sharma$^{2}$ (Member, IEEE)}\\
$^{1}$School of Electrical Sciences, IIT Goa, India			Email: kuntal@iitgoa.ac.in\\ $^{2}$Department of Electronics Engineering, IIT (BHU) Varanasi, India
	Email: sanjeevs.ece@iitbhu.ac.in

}

\maketitle

\begin{abstract}
Next-generation internet
of things (IoT) applications need trillions of  low-powered wireless mobile
devices to connect with each other having ultra-reliability and low-latency.  Non-orthogonal multiple access (NOMA) is a promising technology to  address massive connectivity for 5G and beyond by accommodating several users within the same orthogonal resource block. 
 Therefore, this article explores
hybrid NOMA (HNOMA) for massive 
multiple access  in the uplink scenarios due to its higher spectral efficiency. The HNOMA includes both power domain and code domain NOMA method due to  diverse
channel conditions in practice.
We highlight that polar coded based data transmission can achieve  higher  reliability and lower latency in HNOMA-based wireless networks. 
 Further, at the base station (BS), channel state information (CSI) of each link is not perfectly available or very complex to estimate due to  non-orthogonal links. Therefore, we analyze and  review the  performance of uplink based system  involving HNOMA transmission
in the presence of imperfect CSI.
Furthermore, we summarize some  key technical
challenges as well as  their potential solutions in futuristic IoT applications using HNOMA transmission.
Finally, we offer some design guidelines for HNOMA-based systems using deep learning approach to implement adaptive and efficient wireless networks.
\end{abstract}

\section{Introduction}
Forthcoming society will get  highly benefited  using the data-driven solutions and methods. In the data-driven environment,  trillions of  low-power mobile
devices will connect to communicate and share their decisions. However, connecting trillions of devices together through a central  base station (BS) will be challenging due to limited available orthogonal resources with very low-latency wireless links. 
Therefore, non-orthogonal multiple access (NOMA) and  hybrid NOMA (HNOMA)
transmission techniques can be  suitable methods as compared to orthogonal multiple access  (OMA) techniques  to connect large devices in the internet of things (IoT) scenarios \cite{sharma2019joint, Nam2019, han2019energy}. 
Since the non-orthogonal
transmission techniques enhance spectral efficiency and lower the latency of a wireless system \cite{sharma2019joint}
The HNOMA uses both power domain (PD) and code domain (CD) NOMA  for multiple access. 
Further, HNOMA provides higher spectral efficiency as compared to NOMA and OMA techniques \cite{sharma2019joint}, and we analyze its potential applications and challenges for 6G networks.

IoT applications can be divided into two main categories as massive IoT in which low power devices continuously send their observation to the cloud,  and critical IoT which includes control signaling, healthcare, and remote manufacturing. Some upcoming applications, wireless sensing, and signaling is essential, like in traffic management, self driving vehicles. Therefore, in this article, we discuss  the massive connectivity of low power devices with  ultra-reliability and low-latency based IoT.

\subsection{Smart IoT Network}
\begin{figure*}[htb!]
	\centering \includegraphics[]{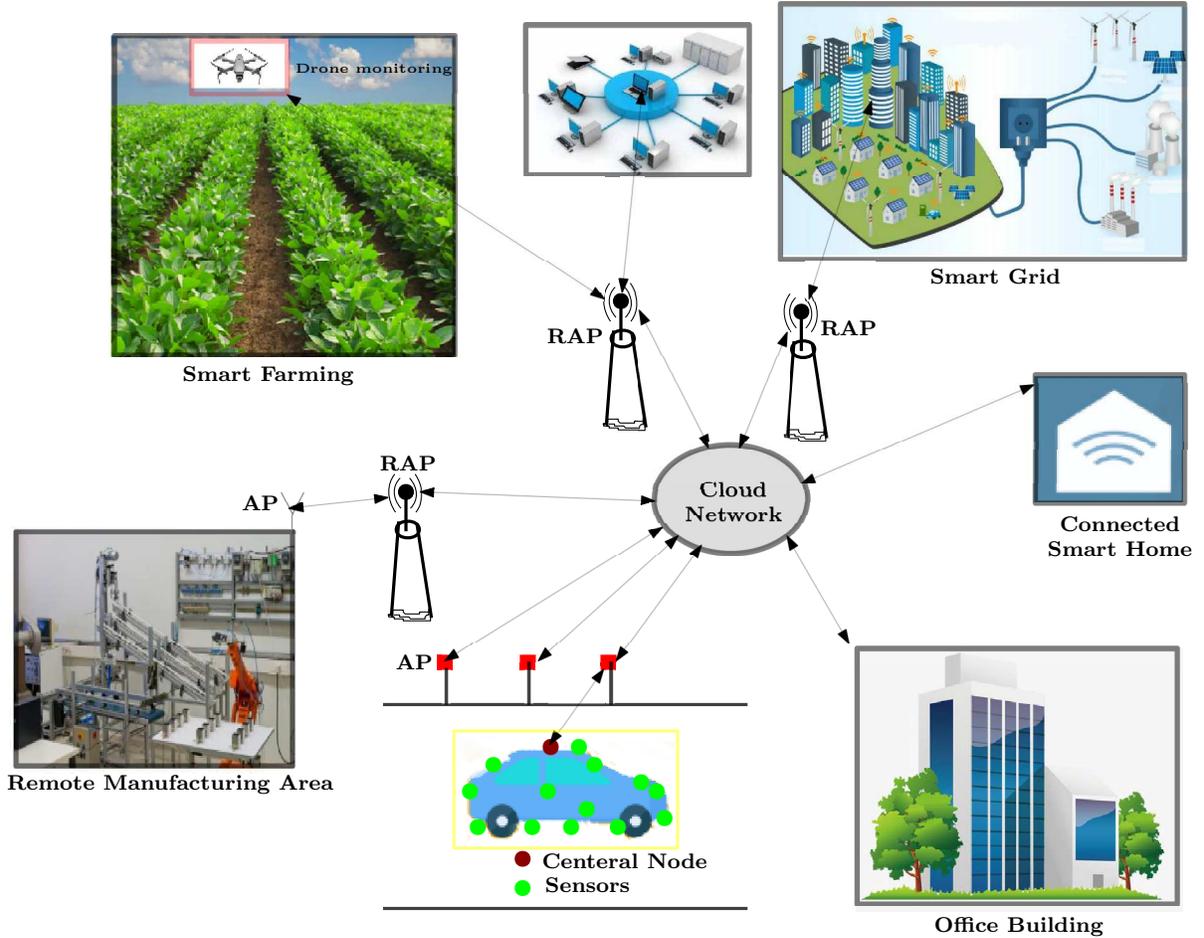}
	\caption {\footnotesize{A view of smart IoT Network. RAP: remote access point, AP: access point.}}	 	
	\label{system_1}
\end{figure*}

 A scenario  of smart IoT applications is shown in Fig. \ref{system_1}. In upcoming  5G and beyond IoT networks, various devices will connect to each other using the virtual cloud networks. For example,
smart farming can benefit from the data-driven design by analyzing the soil and atmospheric conditions of the environment for a particular type of crop. A similar technique can be adopted  for IoT enabled intra-vehicular network (IVN), where  large number of sensors are connected with each other for sharing the vehicle's status information in order to develop a smart vehicular system.  Further, traffic management can be significantly improved by using  sensing and wireless communication methods to  develop smart cities. Therefore, the integration of various areas like industries, offices, power grids, farming, healthcare, etc. using IoT  platform will make society more conformable, flexible and sustainable.

A smart IoT network, heterogeneous infrastructures are connected to a central cloud network, as shown in Fig.
\ref{system_1}. Therefore, each type of infrastructure like smart grid can be optimized  by accessing available data at the central cloud/node. Hence, smart infrastructure  
will give rise to dense/ultra-dense
deployment of access points (AP) in 5G and beyond network.

\subsection{Wireless multiple access techniques}
In order to connect seamlessly trillions low-powered IoT devices, we need higher spectral efficiency and availability of a wide frequency spectrum. Currently, millimeter-wave (mmWave) range of the spectrum is explored for wireless communications \cite{yang20196g}. High bandwidth at mmWave will help to connect large devices together; however, it alone can not be sufficient for IoT applications. Currently, NOMA techniques are preferred over OMA to enhance the wireless systems' spectral efficiency \cite{wang2018non, wan2018non, yang2018non}. In general, either PD or CD NOMA methods are considered in the literature \cite{sharma2019joint}. 
In CD-based NOMA technique includes interleave division multiple access (IDMA), multi-user shared access (MUSA) technique, pattern division multiple access (PDMA), and sparse code multiple access (SCMA) \cite{sharma2019joint}.
In PD NOMA, superposition coding techniques at the transmitter and successive interference cancellation (SIC) at the receiver are used unlike the OMA.
Further, HNOMA, which is an integrated method of PD and CD NOMA, is more suitable and has higher spectral efficiency over conventional  PD or CD NOMA. Therefore, in this article,  we discuss HNOMA for  5G and beyond wireless networks.

\subsubsection{A simple view of wireless connectivity using HNOMA}
\begin{figure}[htb!]
	\centering \includegraphics[]{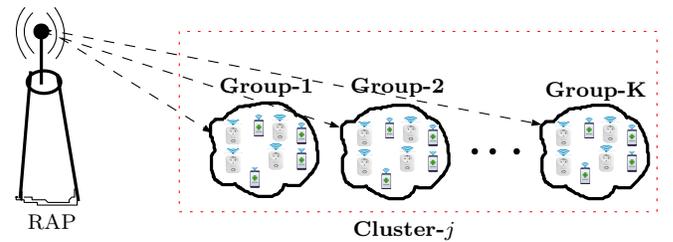}
	\caption {{A sketch for devices connectivity to remote access point using HNOMA. }}	 	
	\label{system_2}
\end{figure}

In Fig. \ref{system_2}, a pictorial representation of the connectivity of low powered devices  to a remote access point (RAP)  is shown.  For easy processing and low latency, the large number of devices are divided into total $J$ clusters in a RAP  area. Further, each cluster has
$K$ multiple  groups based on their distance from the RAP, as shown in Fig. \ref{system_2}.
Therefore, a device ${D}$ in RAP is accessed using cluster and group indices, as ${D}_{j,k}, \ j=1, \cdots J, k=1,\cdots,K$. 
Each  cluster has $Z$ orthogonal (time/frequency/code) resources for communications.
Further, these $Z$ orthogonal resources are reused in other clusters in the system.
Therefore, all the groups  in a cluster use the same  $Z$ orthogonal resources for communications, as shown in Fig. \ref{system_2}. However, $K$  groups are distinguished
using their channel gain at the RAP. Further, devices in each group may have similar or correlated channel impulse response.
Therefore, total $JK$ devices are connected to the RAP using $Z$ orthogonal resources, as depicted in Fig. \ref{system_2}.
In the absence of  strict power control  and sharp frequency cut-off, we assume each cluster may get some interference from other adjacent clusters.  However, we have not considered interference from other adjacent clusters in the system.

\subsubsection{Example: HNOMA
	transmission in V2X communications}
Automotive industries are moving towards connected and autonomous vehicles to  improve traffic congestion, fuel efficiency, safety, etc.  \cite{liang2018toward}.  Therefore, a large number of sensor nodes are required to exchange information. In vehicle-to-everything (V2X) communication, cloud network gather information through several wireless access points such as local area central node, base station, etc.  Therefore, to connect trillions low-powered devices,  high spectral efficiency of network is essential with low latency and efficient signal processing methods. IEEE 802.11p standard is proposed for  V2X Communications to support intelligent transportation  system (ITS) applications. However, still technologies for ITS will have to improve for bounded access channel delay and V2X wireless connections. In V2X, grant free\footnote{User node transmission that does not require  transmit scheduling request i.e. randomly enter or leave the system.} HNOMA with polar coding can be used to achieve low latency and high reliability with high spectral efficiency.

The rest of the article is organized as follows. Given the potential advantages of HNOMA system, in Section~\ref{hybrid_noma}, we illustrate  HNOMA based uplink wireless system for  massive connectivity in IoT. Section~\ref{polar_code} briefly focuses for polar coded HNOMA-based  system.  The impact of channel estimation error in HNOMA  in Section~\ref{estimation}. An example of HNOMA system in the uplink scenario is presented in Section~\ref{uplink}.  After that,  we highlight  some research challenges in upcoming IoT
networks for  heterogeneous applications and HNOMA techniques in Section~\ref{research}.  Then, we discuss and propose a deep learning-based detection and estimation for HNOMA in Section~\ref{ML}.  Section~\ref{conclusions} concludes this article.

\section{HNOMA}\label{hybrid_noma}
Recently,  HNOMA scheme is proposed for multiple access \cite{sharma2019joint, sun2018feasibility}.  The HNOMA scheme using both PD and CD NOMA is more spectral efficient than the PD-NOMA plus OMA-based system \cite{sharma2019joint}. Therefore, 
in this paper, HNOMA uses jointly PD and SCMA-based NOMA.
The  HNOMA uses both power difference and sparsity among the users and orthogonal resources, respectively for transmission. 
Consider the uplink scenario of HNOMA system with two groups only, where Group-$1$ and Group-$2$ have $J_1$ and $J_2$ communicating nodes, respectively.
By considering Group-$1$ and Group-$2$ as the  far and the near groups, respectively, we can get the desired variation in strengths of the  received signals  at the BS.   Group-$2$ (near group)  receives higher  signal power  than the  Group-$1$ in the HNOMA system. 
Therefore, the data for Group-$2$  is decoded first by considering the Group-$1$ signal as an interference using massage passing algorithm (MPA). After decoding Group-$2$, the signal corresponding to  Group-$2$ is subtracted from the received signal at the BS. Then, from the interference-canceled signal,  Group-$1$ symbols are decoded. Therefore, Group-$1$ has full diversity, while Group-$2$ has interference-limited scenarios. Hence, at the BS, both MPA and successive interference cancellation (SIC) based detection is used in HNOMA \cite{sharma2019joint}. In HNOMA, multiple groups can communicate by optimizing transmit power among the different groups and codebook design within a group. Therefore, $K$ groups can communicate to BS in HNOMA, where each group has $J_{k}, k=1,2,...,K$ user nodes. The overloading factor $\lambda$  can be defined as the ratio of the total number $\sum_{k=1}^{K}J_{k}$ of users  to the number  $Z$ of  the orthogonal resources in the system.

Therefore, HNOMA-based system can achieve higher spectral efficiency in the IoT network by marginally increasing the detection complexity. However, the BS or the  central node can support a higher complexity in the system. Further, deep  learning  based information detection can also be used in HNOMA. Deep neural network (DNN) can be trained off-line and is being used for online data detection in wireless communication systems \cite{jiang2016machine}. Furthermore, hybrid non-orthogonal
transmission can also include orthogonal and non-orthognal multiple access schemes for next-generation adaptive networks depending on nodes' data rate constraint.

%

\begin{figure}[h!]
	\centering
	\includegraphics[width=80mm,height=65mm]{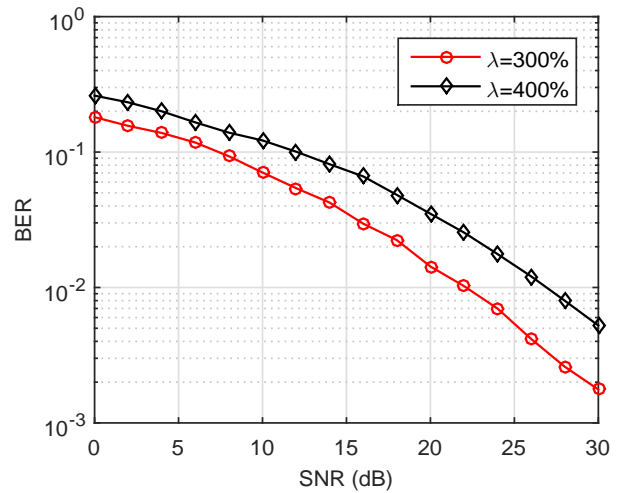}
	\caption {Impact of overloading factor on the hybrid system's SER performance over Rayleigh fading.}
	\label{hybrid}
\end{figure}

The impact of overloading in HNOMA is shown in Fig.~\ref{hybrid}. In each group, the number of user nodes are $J_{1}=J_{2}=6,8$ and orthogonal resources are four, i.e., $Z=4$.  The transmitted signals of the user nodes  experience  independent flat Rayleigh fading. Codebooks in \cite{sharma2019joint} are used by user nodes in Fig.~\ref{hybrid}.
The performance degrades in high overloading scenario ($400\%$) as compared to $300\%$ overloading. However, it is good in  most of  IoT applications. Further, the detection performance of HNOMA-based system can be improved using polar coding techniques.

The HNOMA-based system includes both the SIC and MPA based detection, whose complexity
increases as the number of groups in a cluster and the overloading factor increases. To limit the complexity, the number devices in the IoT networks can be partitioned into small groups and clusters based on their channel gains, as mentioned in Fig. \ref{system_2}. Further, efficient SIC ordering is also essential in HNOMA to reduce the error propagation effect \cite{tse2005fundamentals}.
Further, we introduce SCMA-based NOMA scheme with polar codes to improve the reliability of a wireless network below.

\section{Polar Codes}\label{polar_code}
Erdal Arikan invented a new class of channel codes known as polar codes \cite{arikan_2009}.   These are the first class of codes which can achieve the Shannon capacity in a mathematically provable fashion. These codes are based on the principle of \textit{channel polarization}.  Here, with the help of polar encoding, $n$ synthetic bit channels are manufactured from a given set of $n$ identical channels.  The newly created bit-channels are polarized in the sense that a fraction of them become noise-free  and others become completely noisy channel.  The key aspect here is that the fraction of the noiseless channels is equal to the capacity of the original underlying channel. Therefore, a capacity-achieving coding scheme can be devised  where  the information bits are sent over the noiseless channels and a set of  frozen bits (known to the decoder) are sent over the completely noisy channels.   The decoding is done with the help of successive cancellation (SC) algorithm.  Unlike other modern codes like, Turbo and low-density parity-check (LDPC) codes, the decoding of polar codes is non-iterative.  The performance of the polar codes with SC decoding was not impressive at practical block-lengths.  Then, Tal and Vardy proposed successive cancellation list decoding for polar codes \cite{list}. It has been found that aided by a small cyclic-redundancy check (CRC)  code, the list decoding of polar codes can provide significantly better performance than LDPC and turbo codes when block-length is in the small to medium range.

\begin{figure*}[t]
	\centering	
	\includegraphics[height=5cm, width=15cm]{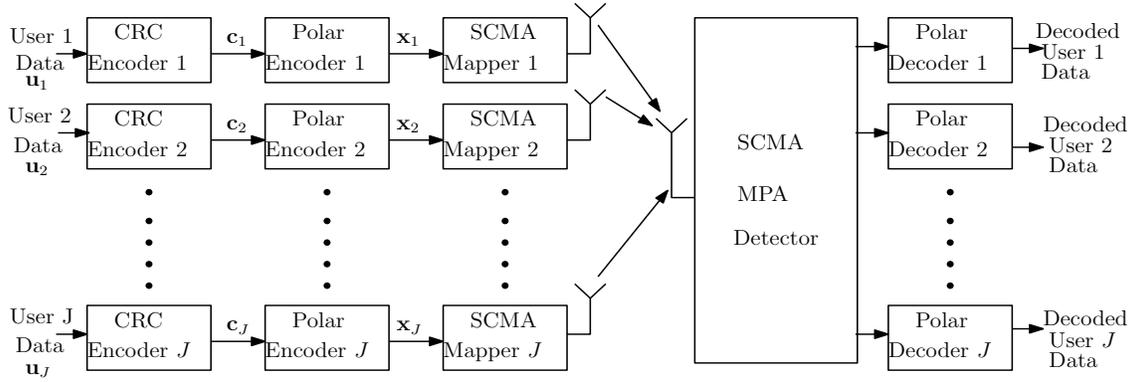}
	\caption{Polar coded SCMA system for uplink.}
	\label{polar_block_diagram}	
\end{figure*}

The performance of an SCMA system can be significantly improved by employing an error correcting scheme.  Here we consider the polar codes as they can provide the best performance amongst all the modern codes.    The block diagram is shown in Fig.~\ref{polar_block_diagram}.  The coding scheme is a concatenated one with a small CRC code of length $n_c$ and polar code of length $n$.  The $j$th user's data are divided into message  frames ${\bf{u}}_j$ of size $k$. Here we have $k=n\times r-n_c$, where $r$ is the rate of the coding scheme. The frame ${\bf{c}}_j$ at the output of the CRC encoder has length $n_c$. The frame ${\bf{c}}_j$ is sent to polar encoder as the input. The output of the polar encoder  is  ${\bf{x}}_j$. Suppose the  SCMA mapper involves $M-$ary constellation.  Then the bits in the frame ${\bf{x}}_j$ are converted into $n_s=\frac{n}{\log_2 M}$ symbols  with each symbol containing $m=\log_2 M$ bits. We have ${\bf{x}}_j=\left({{x}}_j^1, \cdots {{x}}_j^{n_s}\right)$. For every symbol slot ${{s}} \in \left\{1,2, \ldots, n_s\right\}$, the SCMA encoding (mapping) and detection is performed for all the $J$ users.  The SCMA detection is carried out by MPA.  From the MPA detector, the soft values i.e. the probability of an user's data being a particular symbol are considered.  These symbol-wise probability values are converted to bitwise probability values which are  fed to the individual polar decoder.    Every polar decoder employs CRC-aided list decoding algorithm.   The estimated bits across all symbol duration are accumulated and finally  the estimated frame $\hat{{\bf{u}}}_j$ is obtained.

\begin{figure}[h!]
	\centering
	\includegraphics[width=80mm,height=65mm]{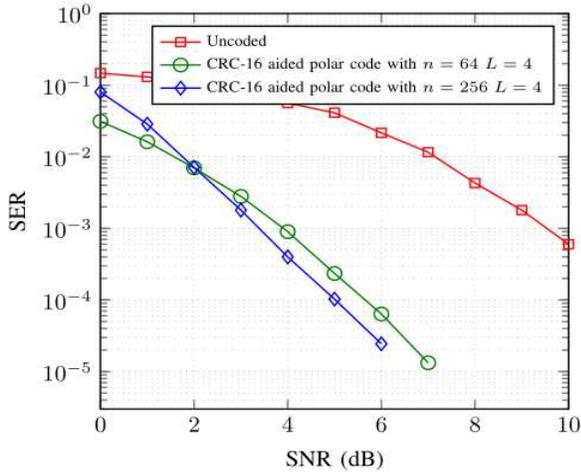}
	\caption {SER performance of various polar coded $6\times4$ SCMA systems.}
	\label{fig1}
\end{figure}

Fig.~\ref{fig1} shows the performance of polar coded SCMA systems with various parameters. In recent communication standards, the  short polar codes are considered for error correction \cite{5G_NR}.  Therefore, we have shown the results for polar codes of block-lengths $n=64$  and $n=256$.  For decoding,  CRC-aided list decoding is used.   The CRC code employed is of length 16 with generator polynomial $g(x) = x^{16} + x^{12} +
x^5 + 1$. This code is known as CRC-16-CCITT.	 The list size of $L=4$ is considered.  The performance can be further improved by considering a higher value of list size.
Observe from Fig.~\ref{fig1} that the with the help of short polar code of length $n=64$, the performance can be drastically improved from that of the uncoded SCMA system.  With a slightly higher value of block-length  $n=256$, the performance can be further enhanced as shown in Fig.~\ref{fig1}.

The selection of the frozen bits known as the \textit{construction} process  is an important step for polar coding. Historically, polar codes have been constructed by Bhattacharyya-parameter-based   algorithm designed for binary erasure channels (BP-BEC). The other popular methods are  Monte-Carlo-based method, density-evolution-based method, Gaussian-approximation (GA) method, Tal and Vardy method, etc.   The features considered for BP-BEC, Monte-Carlo-based method, density-evolution-based method GA and Tal and Vardy method are Bhattacharyya parameter, BER values, probability density function (PDF) of the LLRs, mean of the LLRs   and approximate  PDF of the LLRs respectively.   Here, we consider Monte-Carlo-based method for the construction of polar codes in the multiple-access channel or uplink system.

\section{Imperfect Channel Estimation}\label{estimation}
Symbol detection at the BS requires accurate information of a channel state response of each communicating user node. The received signal is the superposition of each device's data symbol and their channel information.
Therefore, accurate channel estimation at the BS is complex or impossible due to large number of IoT devices with HNOMA schemes.  Since the  BS needs more parameters estimation in the uplink scenario than the downlink. The channel impulse response (CIR) at the BS is expressed as $\textbf{h}=\textbf{h}_{\mathrm{est}}+\textbf{e}$, where
 $\textbf{h}$,  $\textbf{h}_{\mathrm{est}}$, and  $\textbf{e}$ is the actual CIR, estimated CIR and error in channel estimation, respectively. $\textbf{e}$ can be modeled as complex Gaussian random vector with zero mean vector and  the covariance matrix $\sigma_{\mathrm e}^2 \mathbf{I}$. Further, the variance of error depends on the SNR level and is expressed as $\sigma_{\mathrm e}^2=\sigma_{\mathrm h}^2/(1+\rho\text{SNR})$, where $\sigma_{\mathrm h}^2 \mathbf{I}$ is the  covariance matrix of  $\textbf{h}$ and $\rho$
indicates quality of channel estimation. Further, $\rho=\infty$ represents perfect channel estimation in the detection and estimation process. Therefore, the impact of channel estimation should be considered for  massive IoT applications in practice. Some better channel destination algorithms can also be explored in near future for NOMA and HNOMA-based systems.

Further, large channel gain differences  among the groups in a cluster leads to a better channel estimation in HNOMA. However, imperfect channel information also results in residual error in the combined  SIC and  MPA based detection while subtracted the stronger user's signal from the received signal. Therefore, spectrally efficient HNOMA-based systems, channel estimation algorithms and their effect can be explored in near future.


\section{Example of HNOMA-based System  in Uplink Scenario}\label{uplink}
\begin{figure}[h!]
	\centering
	\includegraphics[width=80mm,height=65mm]{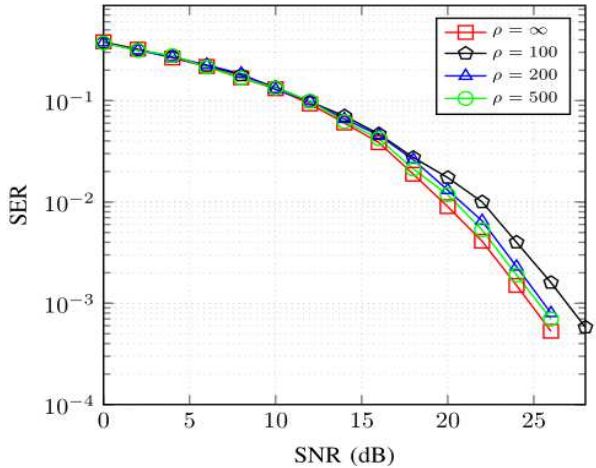}
	\caption {Impact of imperfect channel estimation on SER performance over Rayleigh fading in HNOMA-based systems.}
\label{error}
\end{figure}

We show an example to achieve massive connectivity in IoT using the HNOMA. To reduce the  processing complexity and latency, low power IoT devices are divided into multiple clusters and each cluster has several groups of devices depending on their channel gains. For a single cluster, the received signal  at BS is the superposition of each group's signal and complex additive white Gaussian noise.  
The detection performance in HNOMA depends on the  channel estimation error and noise as shown in Fig. \ref{error}. 
In imperfect channel estimation $\sigma_{\mathrm h}^2=1$ is used.  Overloading factor $\lambda=300\%$ and $Z=4$ are used in Fig. \ref{error}. Imperfect channel estimation limits the performance of HNOMA-based System, as observed in Fig. \ref{error}. However, the impact of channel estimation can be minimized using some coding techniques and complex channel estimation algorithms in the system to achieve  high reliability in IoT networks.

%


\section{Research challenges in upcoming IoT networks}\label{research}

\subsection{Efficient energy enhancement}
In next-generation wireless networks, several kinds of BS like micro, femto  and pico will be  deployed to get ultra-high density of
users. Energy efficiency optimization will provide  green
future cellular networks infrastructure.  In general, using smaller cells, BS power consumption can be reduced. Further, letting BS  go to sleeping mode during light or no traffic load scenarios  result in higher energy efficiency in the network. Therefore, dynamic management of users and BS in the wireless network reduce the power consumption, however, it is a difficult task at large scale level. Therefore, the virtual distribution of resources and their optimization can be a strategic approach to optimize   the energy efficiency of 5G and beyond networks. Further, simultaneous wireless information and power transfer (SWIPT) and wireless power transmission can be used in IoT networks.

\subsection{Asynchronous communications}
Most of the data in upcoming applications will be short, bursty and asynchronous in nature. Therefore, we need some signal processing techniques which will shift the paradigm from current systems to asynchronous and non-orthogonal systems. Further, different mobility level of devices  in the massive IoT will pose severe challenges to achieve lower latency.  Therefore,  detection and estimation  must be robust to asynchronous signaling and must possess the capability to  deal with uncoordinated interference in next-generation wireless networks. Performance of HNOMA and NOMA-based systems can be analyzed in asynchronous and high mobility scenarios for 5G and beyond networks.


\subsection{Optimum devices' grouping and activity detection }
The performance of the HNOMA  mainly depends on the grouping of devices in a cluster. Grouping of devices can be done using their SNR values  at the BS. Similar  SNR devices are partitioned into the same  group in a cluster.
 However, high mobility of  devices and fading can make the user paring difficult in the massive networks to achieve higher spectral efficiency. In literature, some sub-optimal iterative methods are proposed for efficient user paring. Further, using a large data set based on users' history,  efficient user paring can be achieved by employing machine learning approach in complex and large wireless networks.

\subsection{Robust and efficient signal processing}
In some applications like the smart power grid and underwater communications, ambient noise may not be Gaussian in nature i.e., impulsive noise. Therefore, the performance of optimal detection methods (designed using Gaussian noise) can deteriorate significantly. To remove the outliers in large data size using conventional methods like time series before the detection/estimation is a challenging task in massive IoT network due to a higher dimension of the data. Hence, some non-linear signal processing techniques  can be used to efficiently remove/reduce the impact of the impulse noise before the decision in the IoT network. Further, recently deep learning based wireless communication  detection and estimation method can also use to overcome the effect of impulse noise in wireless network. Furthermore, some robust detection/estimation methods are required to overcome the impact of channel estimation error and hardware impairments, in HNOMA-based IoT applications to achieve massive connectivity.

\section{Deep learning based detection}\label{ML}
Humans' daily living  activities will fully integrate to various devices to get a comfortable, sustainable and efficient lifestyle \cite{qolomany2019leveraging}.
Upcoming 5G and beyond networks will be based on data-driven and machine learning (ML) approaches due to extremely high data rate with heterogenous devices and applications. ML-based smart networks will be adaptive, efficient, and predictive in nature.
Therefore, human intervention will be significantly reduced in  smart IoT applications. Further, large IoT network of trillion connected devices will be enhanced using ML techniques.

ML or deep learning will also play a vital role in the health care system to get an easy medical facility in remote areas and early detection of diseases. Further, seamlessly continuous wireless monitoring and assistance will also benefit old age people in  society. Therefore, ML-based solutions give
adaptability, scalability, and flexibility in IoT  applications.

Higher spectral and power-efficient communication systems can be accomplished using deep learning-based detection and estimation methods. The availability of large data size will benefit the deep neural network (DNN) for efficient training and  parameters optimization.
DNN can provide simple and power efficient optimization of  large and complex 5G and beyond networks. Since conventional approaches may be complex and power inefficient in 6G network. For example, recently deep learning based techniques are used in massive MIMO (multiple-input and multiple-output), channel estimation, D2D (device-to-device) for symbol detection, power optimization, and localization in wireless systems \cite{yang20196g}. Another application of deep learning can be used in cognitive radio network for primary user detection and adaptive spectrum management in 6G networks \cite{jiang2016machine}. Further, personal data security and privacy will be  great concerns in next-generation wireless networks. Therefore, ML methods can also enhance  security  by detecting abnormal behaviors and intrusions in network before they happen.

\begin{figure}[h!]
	\centering
	\includegraphics[width=80mm,height=65mm]{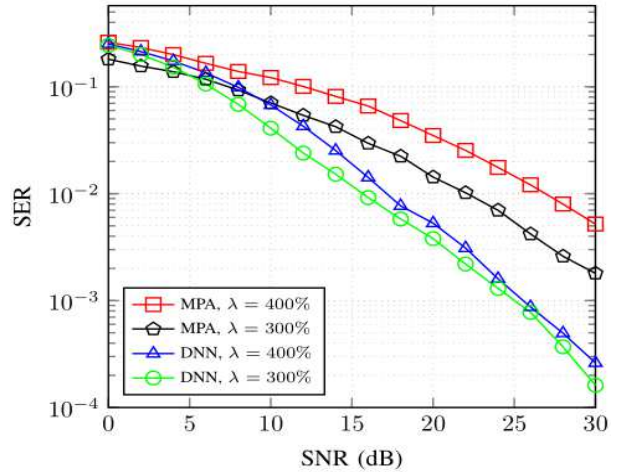}
	\caption {Comparison between DNN and MPA based detection of  hybrid non-orthogonal transmission-based system over Rayleigh fading.}
	\label{dnn}
\end{figure}

Next, we highlight the importance of DNN for symbol detection in HNOMA.
In Fig. \ref{dnn}, the average SER performance of HNOMA using DNN and MPA based detection is shown. DNN outperforms MPA at high SNR as compared to MPA due to effective learning of DNN. MPA performance is also limited by interference in HNOMA system. DNN with input and output, and three hidden layers (with $480$, $10$, $10$ neurons)    trained off-line at $\text{SNR}=30$ dB. DNN also does not need channel information explicitly to decode the users' data symbols, unlike MPA. The training SNR and number of hidden layers are optimized using multiple trials.    Therefore, DNN can help to decode the information in complex and large systems, especially in interference-limited  and imperfect channel scenarios.

%

Further, HNOMA can be analyzed with massive MIMO technology  using conventional and ML-based detection methods to achieve the massive connectivity in low powered IoT network. However, HNOMA with massive MIMO needs some simple and efficient detection methods to deploy this technology in practice. Therefore, the processing power in complex wireless systems  needs to be minimized along  with transmitted power to fulfill the demands of next-generation  green networks. Since at a time, the active devices are very less as compared to the total connected devices in the network. Therefore, ML-based active device detection and their data symbol estimation can also be employed in IoT network for grant-free communication to achieve low latency. .

\section{Conclusion}\label{conclusions}
We have highlighted the importance of HNOMA to connect low power trillion devices in upcoming next-generation IoT applications. The HNOMA can achieve higher spectral efficiency by serving multiple users at a time in the network. Further, the impact of  channel estimation error on the detection performance  is highlighted. The inclusion of  polar codes can  benefit the massive IoT networks to achieve the reliability in adverse and channel mismatch scenarios as highlighted.  Further, some research challenges and their solutions are suggested to fulfill the demand of next-generation IoT networks  in this paper. At last, we have suggested and proposed deep learning-based  smart wireless system design for HNOMA. Since, interference and channel impact can be minimized in HNOMA/NOMA-based systems using deep learning.
Therefore, deep learning and data-driven solutions will play an important role in next-generation wireless networks.

\bibliographystyle{ieeetran}
\footnotesize
\bibliography{references}

%

\end{document}